\documentclass[
reprint,
amsmath,
amssymb,
aps,
prl,
]{revtex4-2}
\usepackage{dcolumn}
\usepackage{bm}
\usepackage{graphicx}
\usepackage{amsmath}
\usepackage{latexsym}
\usepackage{amsfonts}
\usepackage{amssymb}
\usepackage{array}
\usepackage{pifont}
\usepackage{amscd}

\begin{document}

\preprint{APS/123-QED}

\title{Ultraprecise time-difference measurement via enhanced dual pointers with multiple weak interactions}

\author{Yanqiang Guo,$^{1,2}$ Jianchao Zhang,$^{2}$ Jiahui Hou,$^{2}$, Xiaomin Guo,$^{2,}$} 
\altaffiliation{guoxiaomin@tyut.edu.cn}
\author{Liantuan Xiao$^{1,2,}$}
\email{xiaoliantuan@tyut.edu.cn}

\affiliation{$^1$College of Physics, Taiyuan University of Technology, Taiyuan 030024, China\\
$^2$Key Laboratory of Advanced Transducers and Intelligent Control System, Ministry of Education, Taiyuan University of Technology, Taiyuan 030024, China 
}


\date{\today}

\begin{abstract}
Standard weak measurement with an assistant pointer and single weak interaction constrains measurement precision and quantity of interaction parameters, and a compelling characterization of quantum effect featuring weak-value amplification (WVA) remains elusive. Here, we theoretically and experimentally demonstrate an enhanced dual-pointer WVA scheme based on multiple weak interactions and variable spectrum sources. Developing triple weak interactions, momentum P pointer reaches  an optimal time-difference precision of $3.34 \times {10^{-5}}$ as at 6 nm spectral width, and intensity I pointer achieves a displacement resolution of 148.8 fm within 400 kHz linewidth. A quantum effect associated with an anomalous weak value is revealed by an observable violation of a Leggett-Garg inequality. The I-pointer weak value is measured to be 1478 using multiple weak interactions and high signal-to-noise detection, achieving a two-order-of-magnitude WVA enhancement compared to standard weak measurement. Our work opens up a practical avenue for minuscule quantumness measurements in challenging environments.
\end{abstract}

\maketitle


\textit{Introduction.}\textemdash Quantum precision measurement involves determination of interaction parameters and is crucial for elucidating diverse quantum physical effects \cite{Dressel14}. The existing standard model describes the evolution of quantum measurement, allowing the estimation of interaction parameters between a meter state and an observable system \cite{vonNeumann55}. Weak measurement arising from the meter  and system coupling is a significant extension of quantum measurement, and provides a low-disturbance metrological technique to extract a pointer shift \cite{Aharonov88,Lundeen11,Zhang15}. The shift can be effectively amplified with an elaborate postselection, and the amplification protocol related to the weak value is known as weak-value amplification (WVA) \cite{Jozsa07,Jordan14,Ferrie14}. The WVA effect has been actively deployed in precision measurements to enhance tiny physical quantities, such as displacement \cite{Ritchie91,Hosten08}, angular deflection \cite{Dixon09,Binke22}, temperature \cite{Li18,Yanjia19}, frequency \cite{Starling10,Steinmetz22}, magnetic field strength \cite{Yin21}, and enable direct quantum tomography \cite{Xu21,Liang21} as well as weak-measurement microscopy \cite{Liu24}. In standard weak measurement, a single pointer is devoted to extracting the interaction parameters within a single weak interaction (SWI). In particular, time-difference weak measurement employs a single momentum P pointer \cite{Brunner10,XiaoYe13,Bruder13,Zhang16,Fang16,Wang22}, i.e. the shift of the P mean value is used to estimate the optical time difference, revealing that wide spectrum light contributes to the improvement of the measurement sensitivity. High-precision time-difference measurement has been achieved by utilizing a single pointer and a wide spectrum source under single weak interaction \cite{Huang19}. However, the other pointer shift of the spectrum has been overlooked, despite its critical importance in time-difference or multiparameter estimations, particularly as the spectrum width of light source narrows \cite{Qiu17,Zhaoxue18,Luo19}. Furthermore, high precision and multiparameter noninvasive measurement with WVA effect remains a key challenge and is important for advancing applications of quantum metrology and quantum communication.

A prevalent strategy for enhancing measurement precision is leveraging quantum resources \cite{Guo12,Guo18,Guo22}, and continuous-variable squeezing states \cite{Pang15} and entanglement-assisted states \cite{Pang14,Shengshi15,Chen19} are proposed to improve the trade-off between signal intensity and sensitivity for WVA. However, the practical application is limited by the difficulty in preparing premium quantum resources, and pragmatic solutions without using quantum resources is growing in significance. Although entangled photon pairs are utilized in the schemes of sequential \cite{Piacentini16,JiangShan19}, iterative \cite{Kim22} and single click \cite{Enrico21} WVA, they provide valuable insights into mitigating quantum resource constraints and enhancing system robustness, and the practical weak measurements is increasingly becoming a viable reality. More importantly, unequivocal characterization of quantum effect corresponding to anomalous weak value and efficient extraction of the weak value remain to be explored.

In this letter, we present an advanced dual-pointer WVA scheme enhanced by multiple weak interactions (MWI) and variable spectrum sources. The time-difference signal at the ${10^{ - 5}}$ attosecond level and the displacement signal at the hundred femtometer level are measured using P and intensity I pointers, respectively. The shifts of the P and I dual-pointers exhibit direct and inverse proportionality correspondingly to the spectral width of the light source. Furthermore, we showcase a quantum effect of MWI system related to an anomalous weak value by revealing a violation of the Leggett-Garg inequality (LGI). The MWI efficiently extract an anomalous weak value while maintaining a high signal-to-noise ratio unchanged. Our study introduces a practical paradigm for quantum sensing and quantum effect extraction within systems of escalating complexity.

\begin{figure}[htpb] 
\centering
\includegraphics[width=\linewidth]{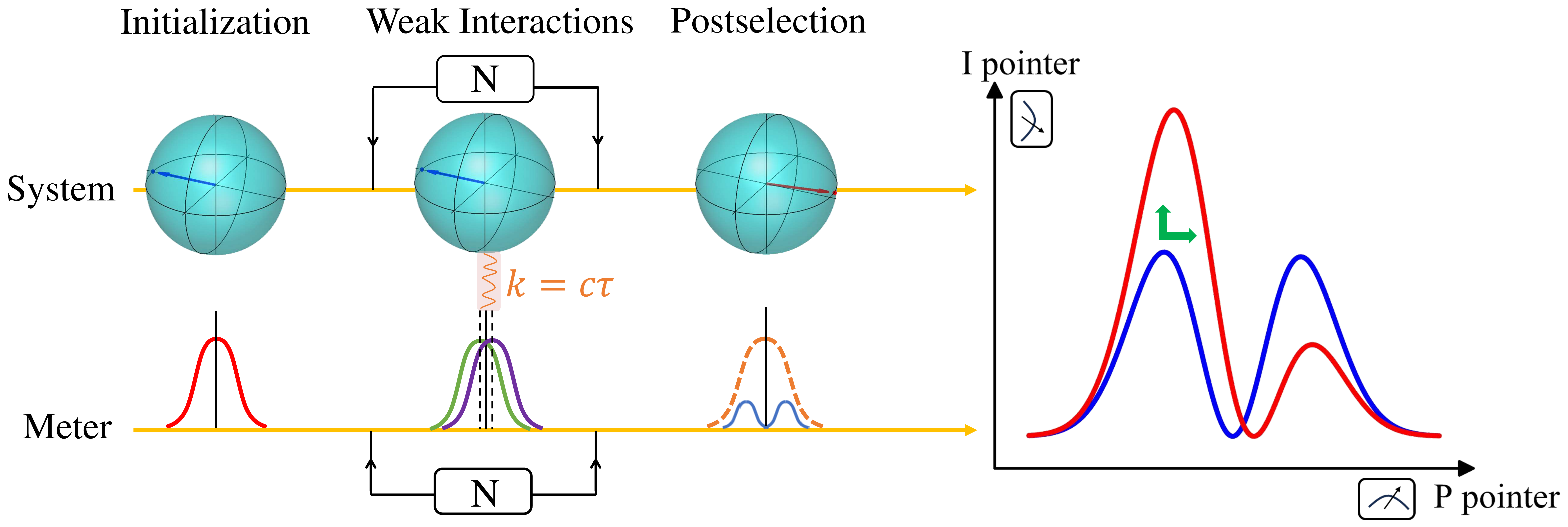}
\caption{Diagram of weak-value amplification (WVA) via enhanced dual pointers with multiple weak interactions (MWI) scheme. The polarization degree of freedom and momentum spectral distribution are set as the observable system and meter state, where photon momentum and total intensity are the P pointer and the I pointer, respectively. The system and the meter state are coupled by a small interaction strength $k = c\tau$ and undergo N multiple interactions during the weak interaction step. Postselection on the system induces a collapse of the meter state into a bimodal distribution., and both the P and I pointers are very sensitive to $k$.}
\label{fig1}
\end{figure}

\textit{Theory.}\textemdash We introduce an enhanced two pointers WVA scheme with the MWI to measure a tiny optical time difference $\tau$ $(\tau  \ll 1)$, and the evolutionary process is shown in Fig.~\ref{fig1}. By referring to the polarization degree of freedom as the physical system, the initial state is given as $\left| {{\psi _i}} \right\rangle  = \left( 1/\sqrt{2}\right) (\left| H \right\rangle  + \left| V \right\rangle )$.The momentum spectral distribution is the meter state, where photon momentum and total intensity are the P pointer and the I pointer, respectively. The photon momentum is ${p_0} = {\omega _0}/c$, where ${{\omega _0}}$ denotes the center frequency corresponding to the center wavelength ${{\lambda _0}}$ of the incident light, and $c$ is the speed of light.

 We begin our analysis with the P pointer. The meter state is initialized to $\int {dp} \left| {\phi (p)} \right\rangle$, which is a Gaussian with center ${{p _0}}$ and standard deviation ${\sigma _p}$. The interaction strength $k = c\tau$ can be obtained from the interaction between the observable system and meter state. For the SWI, the interaction operator is expressed as $\hat U = \exp [ {i\left( {k/2} \right)\hat A\hat P} ]$, where $\hat A = \left| H \right\rangle \left\langle H \right| - \left| V \right\rangle \left\langle V \right|$ is the system observable and $\hat P$ is the momentum operator of the meter state. Given the minimum invasiveness of the weak interaction, multiple interactions between the observable system and meter state carry additional weak interactions information. It is indeed feasible to perform MWI, where the MWI operator is denoted by $\hat \Gamma={\hat U^N} = \exp [ {i\left( {k/2} \right)N\hat A\hat P} ]$. The multiple interactions couple the system to the meter state. Postselection is subsequently performed on the system within the selected projection basis, typically projected onto a state that is nearly orthogonal to the initialization $\left| {{\psi _f}} \right\rangle  = \left( 1/\sqrt{2}\right) \left[ {\exp \left( { - i\rho } \right)\left| H \right\rangle  - \exp \left( {i\rho } \right)\left| V \right\rangle } \right]$, where $\rho$ is the postselection angle. Postselection induces the collapse of the meter state into an unnormalized redistribution as 
\begin{equation}
\label{eq1}
D{(p)_{MWI}} = \frac{{\Phi (p)}}{2}\left[ {1 - \cos \left( {Npk + 2\rho } \right)} \right].
\end{equation}
$\Phi (p) = {\left| {\left\langle {\phi (p)} \right.\left| {\phi (p)} \right\rangle } \right|^2}$ is the initial momentum spatial distribution and $p$ is the eigenvalue of $\hat P$. When $\tau ,\rho  \ll 1$, postselection collapses the meter state distribution into a bimodal profile, maximizing sensitivity to the measured parameter. The observable postselection probability ${P_{MWI}} \approx {\sin ^2}(\rho )$, and the weak value ${\langle {\hat M } \rangle _w}$ for the MWI is
\begin{equation}
\label{eq2}
{\langle {\hat M} \rangle _w} = {\langle {N\hat A} \rangle _w} = \frac{{\left\langle {{\psi _f}} \right|N\hat A\left| {{\psi _i}} \right\rangle }}{{\left\langle {{\psi _f}} \right|\left. {{\psi _i}} \right\rangle }} = iN\cot (\rho ).
\end{equation}
It should be noted that the ${\langle {\hat M } \rangle _w}$ is N times the standard weak value ${\langle {\hat A} \rangle _w} = i\cot (\rho )$. The $k$ can be derived from the shift of the P pointer mean value, which is expressed as follows
\begin{equation}
\label{eq3}
\begin{aligned}
\Delta {p_{MWI}} &= \frac{1}{{2{P_{MWI}}}}\sigma _p^2Nk{e^{ - \sigma _p^2{{\left( {Nk} \right)}^2}}}\sin \left( {Nk{p_0} + 2\rho } \right)\\
 &\approx k\sigma _p^2{\mathop{\rm Im}\nolimits} ( {{{\langle {N\hat A} \rangle }_w}} ).
\end{aligned}
\end{equation}
The approximation is feasible with $k{p_0}/2 \ll \rho  \ll 1$. The $\Delta {p_{MWI}}$ has been augmented by a factor of $N$ compared to that of SWI.

Following the meticulous postselection amidst the multiple system-meter interactions, we can observe the horizontal shift $\Delta {p_{MWI}}$ of the mean value of photon momentum P pointer and the vertical shift $\Delta {\ell _{MWI}}$ of the total light intensity I pointer in the collapsed meter state. When the incident momentum spectrum distribution is wide, the $\Delta {p_{MWI}}$ is more pronounced relative to the $\Delta {\ell _{MWI}}$. However, as the incident light spectrum narrows, the shift of the I pointer becomes more prominent. Especially when a single-mode coherent light is incident, there is virtually no horizontal shift in the meter state, as the weak interactions are entirely converted into the vertical intensity shift. It is worth noting that high-precision measurement across varying spectral incident states cannot be achieved using a single pointer.

Since the I pointer is reflected in the intensity, let the initial intensity without postselection be ${I_{init}}$. Following the MWI and the postselection $\left| {{\psi _f}} \right\rangle$, the intensity detected evolves as ${I_{MWI}} = {I_{init}}{P_{{{MWI}}}}$. The shift of the I pointer with MWI is denoted as
\begin{equation}
\label{eq4}
\begin{aligned}
\Delta {\ell _{MWI}} &= \frac{{\Delta {I_{MWI}}}}{{I_{MWI}^{k = 0}}} = \frac{{{I_{MWI}} - I_{MWI}^{k = 0}}}{{I_{MWI}^{k = 0}}}\\
 &\approx {e^{ - \sigma _p^2{{\left( {Nk} \right)}^2}}}{p_0}k{\mathop{\rm Im}\nolimits} ( {{{\langle {N\hat A} \rangle }_w}} ).
\end{aligned}
\end{equation}
Compared to that of SWI, the $\Delta {\ell _{MWI}}$ has been enlarged by a factor of $N$ (see detail in Supplemental Material, Sec.1 \cite{Supple}).

The precision $\delta k$ is given by $\delta k = \delta m/(\frac{{\partial S}}{{\partial k}})$, where $\delta m$ is the resolution of the P pointer determined by the instrument measuring the momentum spectral distribution, or the uncertainty of the I pointer. $S$ corresponds to $\Delta {p_{MWI}}$ and $\Delta {I_{MWI}}$, and $\partial S/\partial k$ denotes the shift rate. Since the shift of MWI is $N$ times that of SWI, the precision is improved by $1/N$ in the MWI scenario.

\begin{figure}[htpb]
\centering
\includegraphics[width=\linewidth]{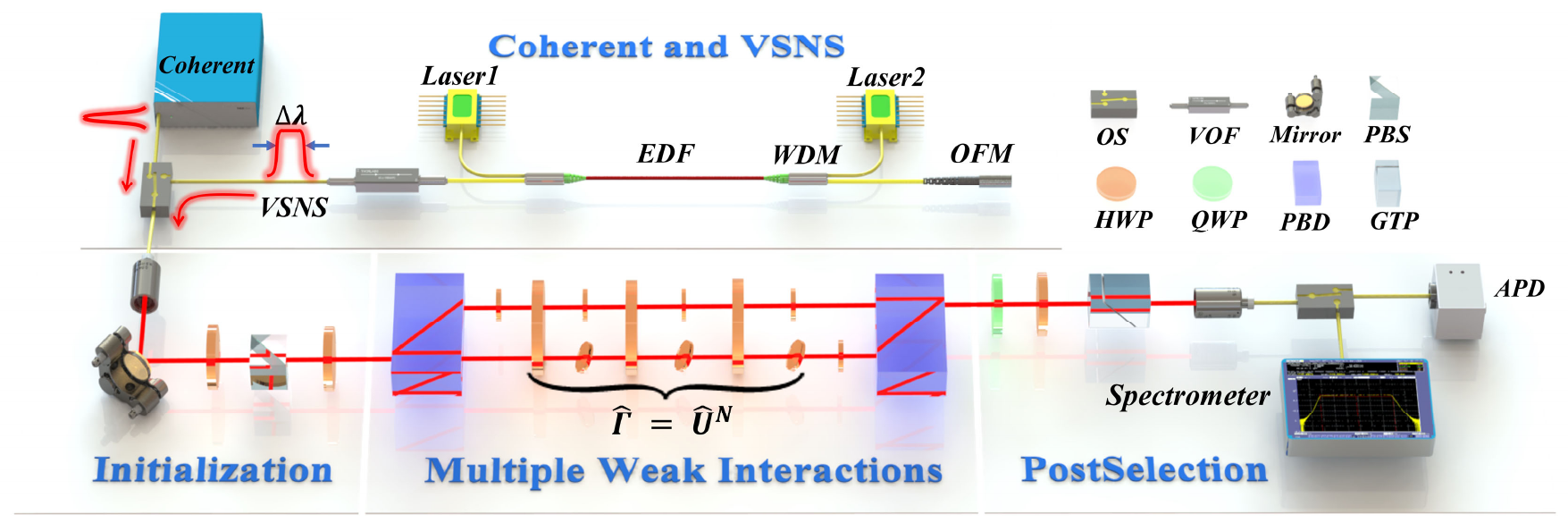}
\caption{Experimental schematic of ultraprecise time-difference $\tau$ measurement via dual-pointers with MWI. VSNS: variable spectrum noise source; EDF: erbium-doped fiber; VOF: variable optical filter; HWP: half wave plate; PBS: polarization beam splitter; PBD: polarization beam displacer; QWP: quarter wave plate; GTP:  Glan–Taylor polarizer. The final meter state through an optical switch (OS) either into the spectrometer for measuring the shift of the P pointer or into the avalanche photodiode (APD) for measuring the shift of the I pointer.}
\label{fig2}
\end{figure}

\textit{Experiment.}\textemdash To experimentally verify the superiority of dual pointers and the enhancement of the MWI in measuring an ultrasmall $\tau$, we develop a practical optical system depicted in Fig.~\ref{fig2}. A variable spectrum noise source (VSNS) is employed instead of a wide-spectrum Gaussian source commonly used in previous studies. The erbium-doped fiber (EDF) is subjected to bidirectional pumping by two 980 nm high-power lasers, facilitating the emission of amplified light in both forward and backward directions. The backward amplified light is reflected by an optical fiber mirror (OFM) to repump the EDF and merges with the previous forward amplified light through two wavelength division multiplexers (WDMs). The final combined light passes through a variable optical filter (VOF) to select the spectrum width. The VSNS output spectrum is flat on top, with a central wavelength of 1550 nm (see detail in Supplemental Material, Sec.2 \cite{Supple}). We additionally employ a coherent source, with a center wavelength of 1550 nm and a linewidth of 400 kHz. The polarization degree of freedom serves as the observable system, and the momentum spectral distribution $\left| {\zeta (p)} \right\rangle$ serves as the meter state. 

In the initialization module of Fig.~\ref{fig2}, the preselected state is prepared to $\left| {{\psi _i}} \right\rangle$ using two half wave plates (HWPs) and a polarization beam splitter (PBS). In the MWI module, the preselected input light traverses a polarization beam displacer (PBD), and reaches the first HWP with its optical axis set at $45^\circ$. A single HWP with its optical axis set at $0^\circ$ accompanied by two HWPs with their optical axes set at $90^\circ$ constitutes a set of weak interactions, with three such sets facilitating triple weak interactions. The HWP of the lower path is tilted at a certain angle $\theta$ to introduce the interaction strength $k=c\tau$. The ultrasmall time difference $\tau$ to be measured as a function of the angle $\theta$ is
\begin{equation}
\label{eq5}
\tau  = \frac{\lambda }{{2\pi c}}(\frac{{\pi L}}{{{L_0}}} - \pi ) = \frac{\lambda }{{2c}}(\frac{1}{{\sqrt {1 - {{\sin }^2}\theta /{n^2}} }} - 1),
\end{equation}
where $n = 1.54$ is the HWP refractive index at $\lambda=1550$ nm.

Due to the realistic experiment condition, it is unavoidable to introduce an initial phase difference $\gamma p$ between the two polarization components $\left| H \right\rangle$ and $\left| V \right\rangle$ in the interferometric optical path. Then the interaction system-meter state is rewritten as$\int {dp}\left( 1/\sqrt{2}\right) \left[ {{e^{i{p(}N{k+\gamma )/2}}}\left\vert H\right\rangle +{e^{-i{p(}N{k+\gamma )/2}}}\left\vert V\right\rangle }\right] {\left\vert {\zeta (p)}\right\rangle }$. The initial phase difference serves as a kind of modulation so that the P pointer shift is modulated to the position of the initial phase difference rather than to the position of the true optical zero-time difference. The advantage of utilizing imaginary weak values has been elucidated \cite{Brunner10,XiaoYe13,Bruder13,Zhang16,Fang16,Wang22}. To achieve an amplified detection of $\tau$, we employ a combination of a quarter wave plate (QWP), a HWP and a Glan–Taylor polarizer (GTP) for postselection. In our experiment, accounting for the practical modulation effects of initial phase difference, the impact of the QWP can be neglected. In the postselection, the polarization mode of the system is projected to $\left| {{\psi _f}} \right\rangle$, resulting in the collapse of the meter state. The recombination spectral distribution following the postselection is rewritten as $\frac{{\Omega (p)}}{2}\{ 1 - \cos [p(Nk + \gamma ) + 2\rho ]\}$. $\Omega (p) = {\left| {\left\langle {\zeta (p)} \right.\left| {\zeta (p)} \right\rangle } \right|^2}$ represents the initial momentum spectrum distribution in our experiment. From the relationship of $p = 2\pi /\lambda$, the $\Delta p$ transforms into the shift of the wavelength mean value $\Delta \lambda$. The final meter state is coupled through an optical switch (OS) into a spectrometer and an avalanche photodiode (APD) for measuring the shift of the P pointer and I pointer, respectively (see detail in Supplemental Material, Sec.2 \cite{Supple}). Our spectrometer (APEX AP2060A) has a resolution of 0.04 pm, and the APD has a conversion gain of $3.14 \times {10^6}$ V/W with a bandwidth of 10 MHz.

\begin{figure}[htpb]
\centering
\includegraphics[width=\linewidth]{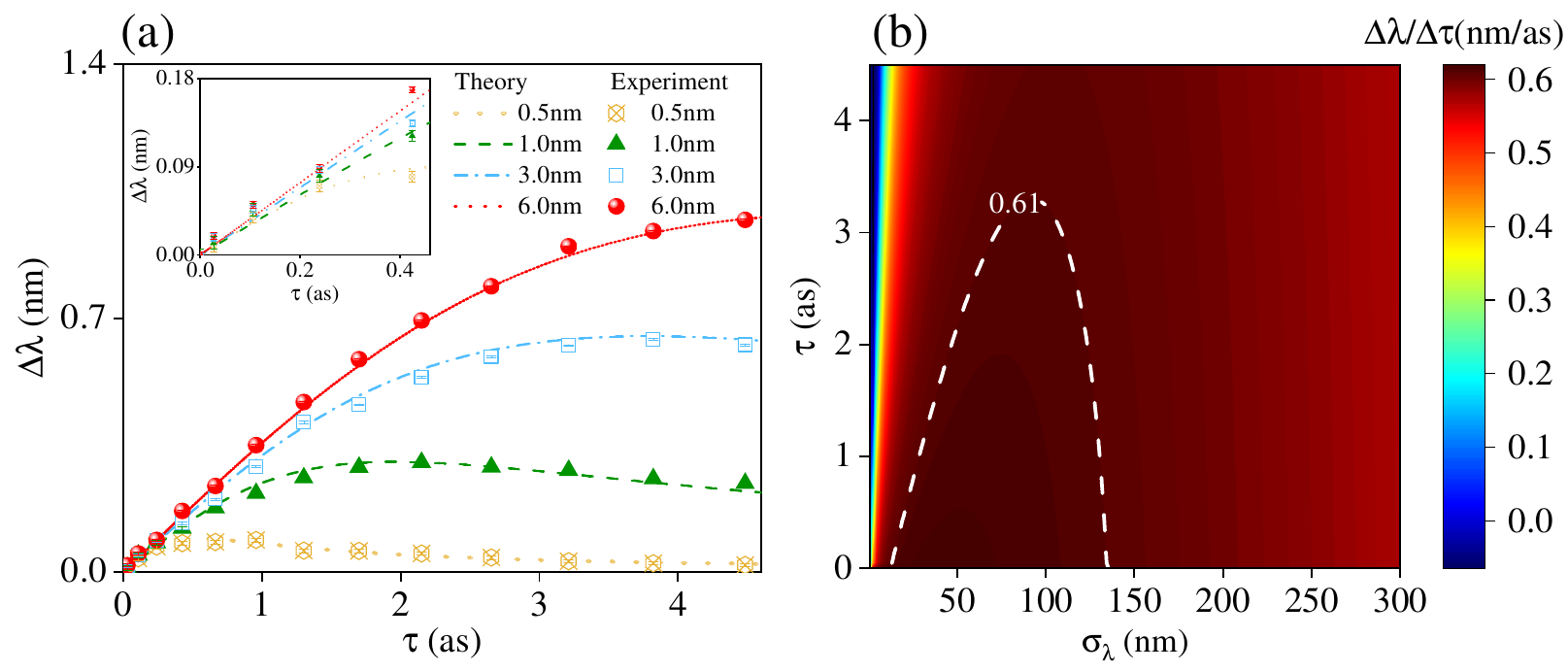}
\caption{(a) $\Delta \lambda$ versus $\tau$ for various spectral widths of VSNS (${\sigma _\lambda }=0.5$ nm, 1.0 nm, 3.0 nm and 6.0 nm). The dots are experimental results and the curves denote theoretical fittings. The error bars for each experimental dot are the standard deviation of twenty repetitive experiments, which can be seen in the inset. (b) Map of shift rate $\Delta \lambda$ with $\tau$ and $\sigma _\lambda$. The dashed region depicts the  maximum shift rate of 0.61 nm/as.}
\label{fig3}
\end{figure}

\begin{figure}[htpb]
\includegraphics[width=\linewidth]{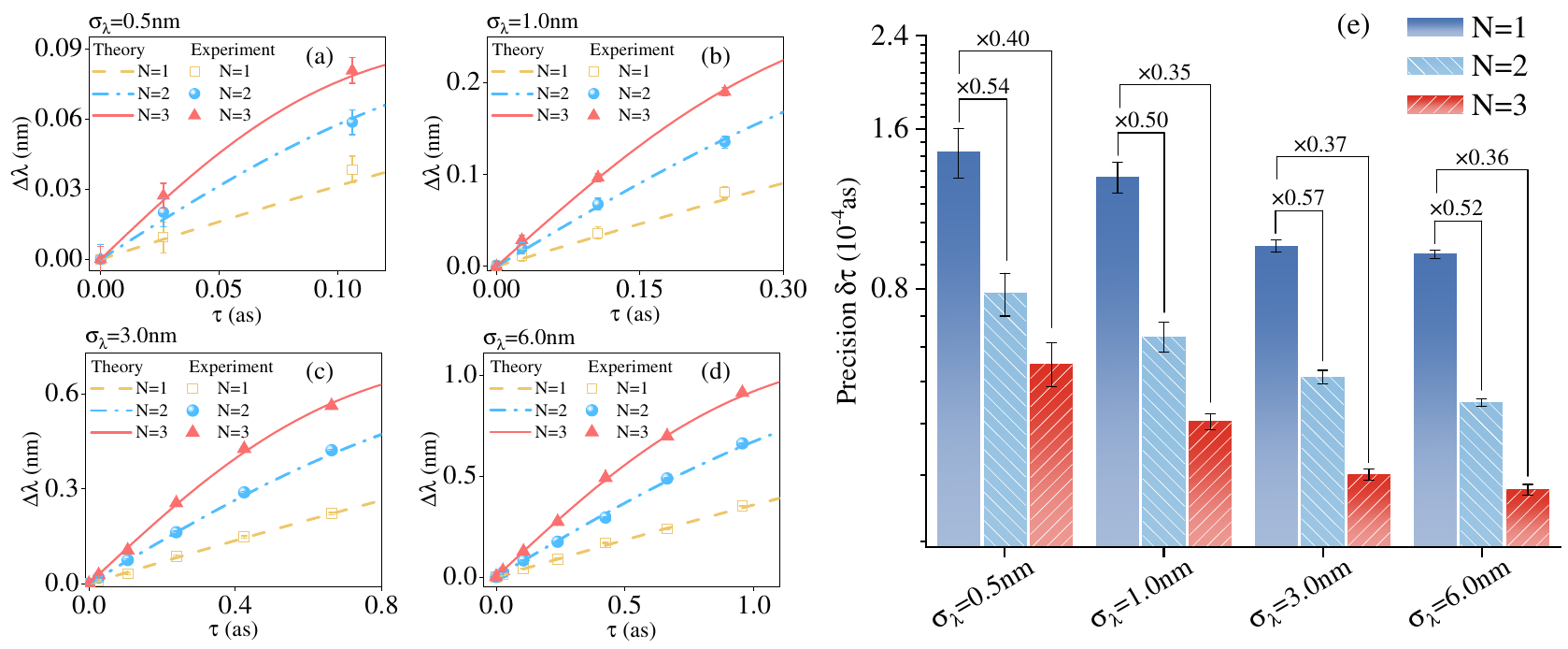}
\caption{(a)-(d) $\Delta \lambda$ versus $\tau$ based on the MWI for various spectral widths of VSNS. The dots and lines respectively represent the experimental and theoretical results. (e) The detection precision $\delta \tau$ with the MWI. The error bars are the standard deviation of twenty repetitive experiments in all cases.}
\label{fig4}
\end{figure}

\textit{Experimental results.}\textemdash In Fig.~\ref{fig3}, we first show the results of the time difference $tau$ with respect to the change in $\Delta \lambda$ under the SWI. Figure~\ref{fig3}(a) illustrates the measurements for the VSNS spectral widths of ${\sigma _\lambda }=0.5$ nm, 1.0 nm, 3.0 nm and 6.0 nm, when the postselection angle $\rho$ is set to 0.002 rad. The optimal fit between theory and experiment is achieved at an initial phase difference of $\gamma p \sim 1.9\pi$. From the $\delta k = \delta m/(\frac{{\partial S}}{{\partial k}})$, the shift rates $\Delta \lambda /\Delta \tau$ in the linear region for the four VSNS widths are measured to be 0.27 nm/as, 0.31 nm/as, 0.41 nm/as and 0.43 nm/as, respectively. The corresponding precisions are $1.45 \times {10^{ - 4}}$ as, $1.30\times{10^{-4}}$ as, $9.62 \times {10^{ - 5}}$ as and $9.30 \times {10^{ - 5}}$ as, respectively. But for wider spectral widths, the shift rates $\Delta \lambda$ does not always increase with the spectral width. As shown in Fig.~\ref{fig3}(b), we analyze the shift rates under the VSNS incidence within a spectral width range of 300 nm. In the case of the experiment initial phase difference, it is observed that $\Delta \lambda$ reaches the maximum 0.61 nm/as within the optimal spectral range 12\textendash135 nm, and as the spectral width increases further, the shift rate decreases instead. 
Then we develop the MWI within the linear regime to enhance the time difference measurements. Figure~\ref{fig4}(a)\textendash\ref{fig4}(d) reveals that the shifts $\Delta \lambda$ resulting from the MWI at $N=2$ and $N=3$ are respectively twice and three times the SWI at $N=1$. But as the $\tau$ increases, especially when it deviates from the linear range, the amplification factor is not exactly $N$ times greater as the approximation condition ($k{p_0}/2 \ll \rho  \ll 1$) does not  adequately holdsatisfy well. As expected, Figure~\ref{fig4}(e) shows that the precision $\delta \tau$ with the MWI has improved by nearly $1/N$ compared to the SWI. The optimal precision is $3.34 \times {10^{ - 5}}$ as, achieved with an incident 6.0 nm VSNS under triple weak interactions.

\begin{figure}[htpb]
\includegraphics[width=\linewidth]{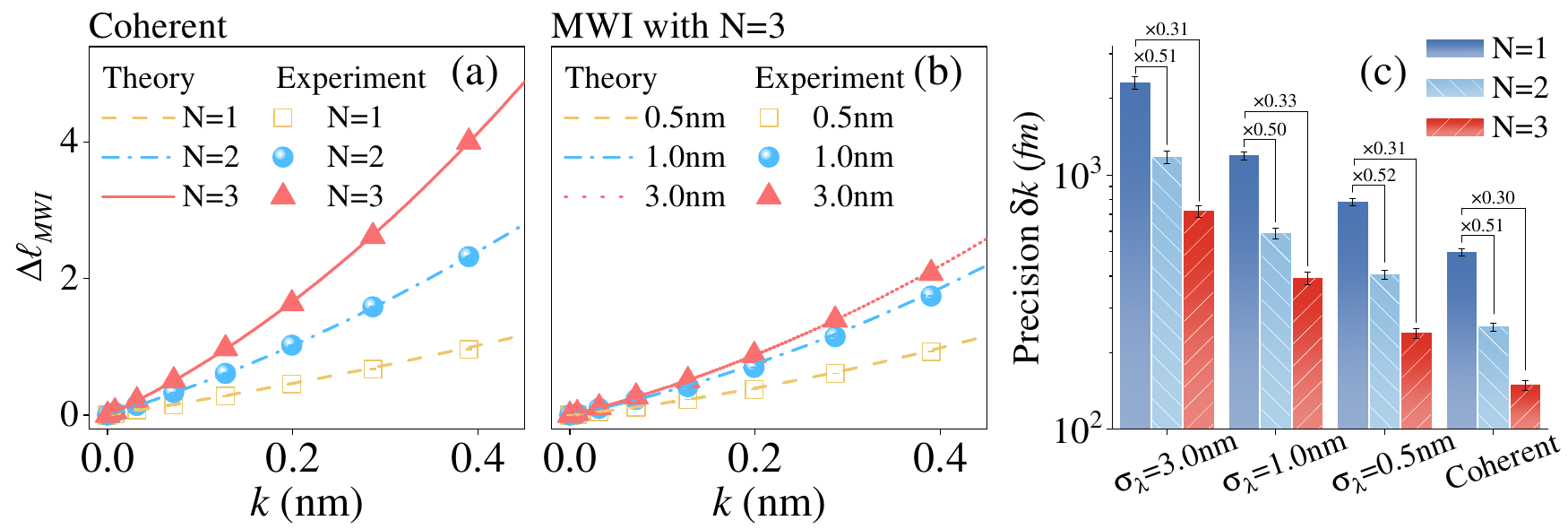}
\caption{(a) and (b) displacement difference $k$ versus intensity shift $\Delta {\ell _{MWI}}$ with the MWI for coherent source and VSNS, respectively. The dots and curves are experimental and theoretical results. (c) Measurement precision $\delta k \propto 1/N$ with the MWI. The error bars represent the standard deviation of twenty repeated experiments in all cases.}
\label{fig5}
\end{figure}

To reveal the role of the other I pointer, we also perform high-precision measurements of the displacement difference $k$ using high signal-to-noise detection and a narrow-linewidth coherent light. According to Eq.~(\ref{eq4}), the shift of I pointer is inversely proportional to the spectral width of incident light, contrasting with the P pointer shift. For an incident single-mode coherent source, the signal spectral distribution is overwhelmed by the electrical noise of -75 dBm, rendering its bimodal distribution indistinguishable. However, a substantial signal-to-noise ratio (SNR) of up to 17.5 dB is achieved using the APD (see detail in Supplemental Material, Sec.2 \cite{Supple}) to measure the postselected state of the narrow-linewidth coherent light. Figure~\ref{fig5} illustrates the measured results of the displacement difference $k$ by implementing the MWI for $\rho=0.002$ rad. Figure~\ref{fig5}(a) and \ref{fig5}(b) show the shift $\Delta {\ell _{MWI}}$ of the I pointer detected by the APD varying with the $k$. As expected by Eq.~(\ref{eq4}), Fig.~\ref{fig5}(a) shows that for $N=2$ and $N=3$, the MWI results in the shift $\Delta {\ell _{MWI}}$ increases of twice and three times that of the SWI, respectively. Figure.~\ref{fig5}(b) demonstrates an inverse relationship between the $\Delta {\ell _{MWI}}$ and the spectral width. Figure.~\ref{fig5}(c) reveals that the enhancement effect of the MWI leads to the precision $\delta k$ improvement of $1/N$. Since the uncertainty of the intensity $\delta I$ for the coherent source is less than 0.044 mV, the optimal precision $\delta k$ of 148.8 fm is achieved with a source linewidth of 400 kHz under $N=3$.

To explore quantum effect of the MWI measurement, we experimentally extract an anomalous weak value and demonstrate a violation of the Leggett-Garg inequality (LGI) based on the WVA scheme. The standard LGI is \cite{Williams08,Pan20}
\begin{equation}
\label{eq6}
{K_3} = {s_1}{s_2}\langle {{{\hat S}_1}{{\hat S}_2}} \rangle  + {s_2}{s_3}\langle {{{\hat S}_2}{{\hat S}_3}} \rangle  - {s_1}{s_3}\langle {{{\hat S}_1}{{\hat S}_3}} \rangle  \le 1,
\end{equation} 
where ${s_i} =  \pm 1$ is the eigenvalue of the corresponding operator ${{{\hat S}_i}}$, denoting initialization ${{{\hat S}_1}}$, weak interaction ${{{\hat S}_2}}$, and postselection ${{{\hat S}_3}}$ of weak measurement. Due to the weak value of Pauli observable is independent of the coupling strength \cite{Liang21,JiangShan19}, it circumvents the necessity for the noninvasive measurability. In our weak measurement, the imaginary part of weak value ${{\mathop{\rm Im}\nolimits} ( {{{\langle {N{{\hat S}_2}} \rangle }_w}} )}$ is determined by the weak interactions and the postselection angle $\rho$, and the ${K_{31}}$ inequality, one of the four LGIs \cite{Pan20}, can be expressed as 
\begin{equation}
\label{eq7}
\begin{aligned}
{({K_{31}})_{MWI}} &= 2{P_{MWI}}[ {1 - {\mathop{\rm Im}\nolimits} ( {{{\langle {N{{\hat S}_2}} \rangle }_w}} )} ]\\
 &\approx 2{\sin ^2}\rho \left( {1 - N\cot \rho } \right),
\end{aligned}
\end{equation}where the postselection probability ${P_{MWI}} \approx {\sin ^2}\rho$ and the ${\langle {N{{\hat S}_2}} \rangle _w} = iN\cot \rho$ are obtained from Eq.~(\ref{eq4}). The violation of the LGI inequality occurs when ${({K_{31}})_{MWI}} < 0$, which is referred to as the quantum effect regime. In this regime, weak values exceeding 1 indicate the presence of anomalous weak values.

\begin{figure}[htpb]
\includegraphics[width=\linewidth]{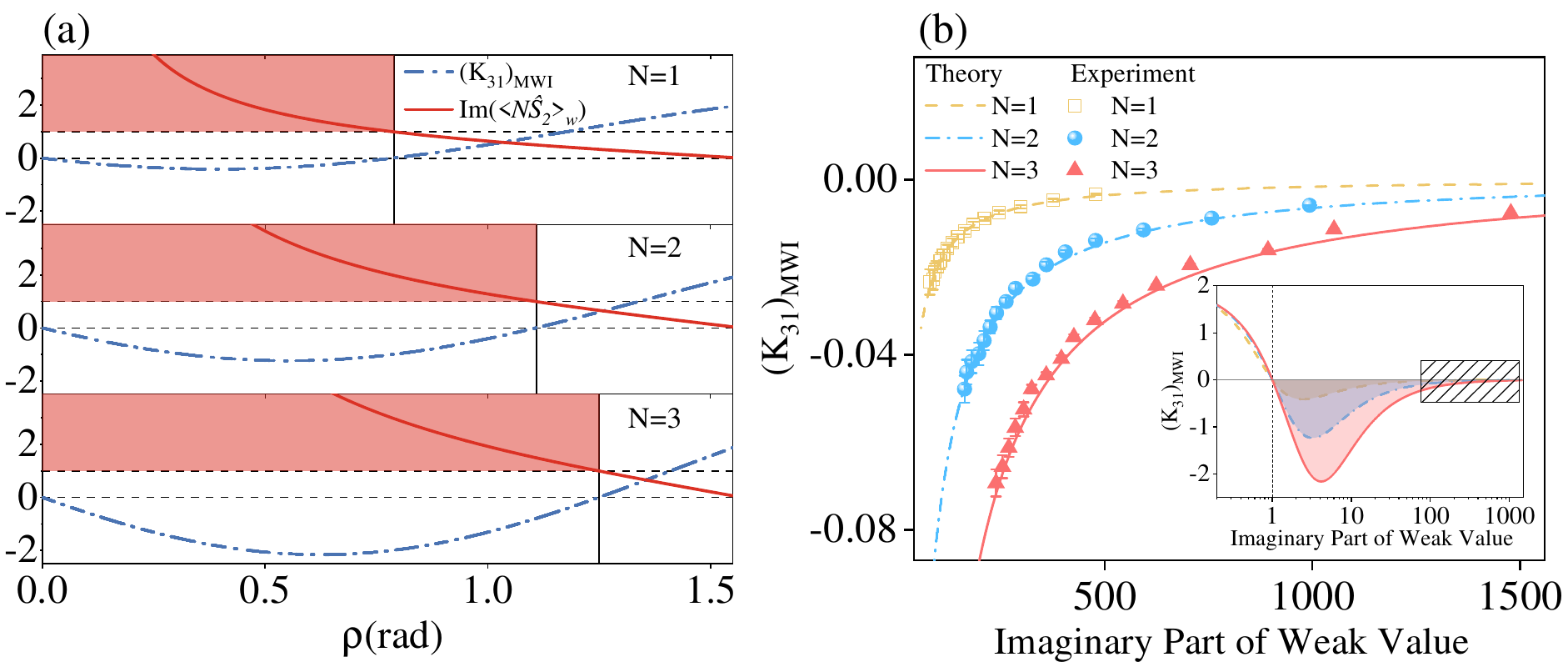}
\caption{(a) Quantum value ${({K_{31}})_{MWI}}$ and weak value ${{\mathop{\rm Im}\nolimits} ( {{{\langle {N{{\hat S}_2}} \rangle }_w}} )}$ as a function of postselection angle $\rho$ for MWI. The shaded portions indicate the quantum effect regions that violate the LGI. (b) ${({K_{31}})_{MWI}}$ versus ${{\mathop{\rm Im}\nolimits} ( {{{\langle {N{{\hat S}_2}} \rangle }_w}} )}$ under MWI. The dots and curves are the experimental and theoretical results. The error bars are the standard deviation of twenty repetitive experiments in all cases. The illustration is the theoretical prediction for a larger range of ${{\mathop{\rm Im}\nolimits} ( {{{\langle {N{{\hat S}_2}} \rangle }_w}} )}$.}
\label{fig6}
\end{figure}

In Fig.~\ref{fig6}, theoretical and experimental results showcase the violation of LGI inequalities and the impact of the anomalous weak values on the quantum values ${({K_{31}})_{MWI}}$. The negative quantum values ${({K_{31}})_{MWI}}$ reveal the quantum effect of the weak measurement, corresponding to the emergence of anomalous weak values. The theoretical region of quantum effect expands as the interaction $N$ increases, as shown in the shaded area of Fig.~\ref{fig6}(a). Figure~\ref{fig6}(b) shows the experimental measurement of the ${({K_{31}})_{MWI}}$ versus ${{\mathop{\rm Im}\nolimits} ( {{{\langle {N{{\hat S}_2}} \rangle }_w}} )}$ for postselection angle 0.002\textendash0.012 rad and $N=1$, 2, 3, respectively. ${{\mathop{\rm Im}\nolimits} ( {{{\langle {N{{\hat S}_2}} \rangle }_w}} )}$ is proportional to the weak interaction $N$, and implementing the triple weak interaction $N = 3$ and high SNR = 32.6 dB, an anomalous weak value of 238 is obtained for the postselection angle $\rho  = 0.0124$ rad. As the $\rho$ decreases, a more significant anomalous weak value of 1478 is observed with the SNR maintaining above 15 dB (see detail in Supplemental Material, Sec.2 \cite{Supple}). Such anomalous weak value with the triple weak interactions reveals an intriguing phenomenon, demonstrating a deeper violation of the LGI and the quantum effect enhancement.

\textit{Conclusion.}\textemdash In our work, we introduce and experimentally validate a dual-pointer WVA scheme based on the MWI, achieving a time-difference precision of $3.34 \times {10^{ -5}}$ as for the P pointer and a displacement resolution of 148.8 fm for the I pointer. For incident sources with diverse spectral widths, the P pointer shift is directly proportional to the spectral width, while the I pointer exhibits an inverse relationship. Furthermore, the violation of the LGI related to the anomalous weak values unveils the quantum effect in the system. Without additional quantum resources, an anomalous weak value up to 1478 is extracted through high SNR detection coupled with triple weak interactions, which is two orders of magnitude than standard weak measurements. The advancement of the dual-pointer WVA scheme with MWI contributes to the applications of precision metrology, quantum sensing and quantum tomography.

\begin{acknowledgments}
This work was supported by the National Key Research and Development Program of China (Grant No. 2022YFA1404201), National Natural Science Foundation of China (Grants No. 62075154, No. 62175176, and No. U23A20380).
\end{acknowledgments}

\nocite{*}

\bibliography{reference}

\pagebreak 
\clearpage
\widetext
\renewcommand{\thefigure}{S\arabic{figure}}
\setcounter{figure}{0}
\renewcommand{\thetable}{S\arabic{table}}
\setcounter{table}{0}
\renewcommand{\theequation}{S\arabic{equation}}
\setcounter{equation}{0}

\section{Supplementary Materials to: Ultraprecise time-difference measurement via enhanced dual pointers with multiple weak interactions}
\subsection*{Yanqiang Guo,$^{1,2}$ Jianchao Zhang,$^{2}$ Jiahui Hou,$^{2}$ Xiaomin Guo,$^{2,*}$ Liantuan Xiao$^{1,2,\dag}$}
\subsection*{$^1$College of Physics, Taiyuan University of Technology, Taiyuan 030024, China\\
$^2$Key Laboratory of Advanced Transducers and Intelligent Control System, Ministry of Education, Taiyuan University of Technology, Taiyuan 030024, China\\
$^*$guoxiaomin@tyut.edu.cn\\
$^\dag$xiaoliantuan@tyut.edu.cn}

\section{\label{sec1}Supplementary Section 1: Enhanced dual-pointer weak-value amplification based on multiple weak interactions}
As depicted in Fig.~\ref{fig1} of the main text, we start from the single weak interaction (SWI), the interaction operator is expressed as $\hat U = \exp \left[ {i\left( {k/2} \right)\hat A\hat P} \right]$. Here the unknown parameter $k=c\tau $, $\hat A = \left| H \right\rangle \left\langle H \right| - \left| V \right\rangle \left\langle V \right|$ is the system observable and $\hat P$ is the momentum operator of the meter state. For the preselected system $\left| {{\psi _i}} \right\rangle  = \frac{1}{{\sqrt 2 }}(\left| H \right\rangle  + \left| V \right\rangle )$ and the meter $\int {dp} \left| {\phi (p)} \right\rangle$ , SWI evolves the coupling system-meter state as [1-4]
\begin{equation}
\label{eqS1}
\tag{S1}
\left| {{\Psi _{SWI}}} \right\rangle  = \hat U\left| {{\psi _i}} \right\rangle \int {dp} \left| {\phi (p)} \right\rangle
 = \int {dp\frac{1}{{\sqrt 2 }}({e^{i\frac{{kp}}{2}}}\left| H \right\rangle  + {e^{ - i\frac{{kp}}{2}}}\left| V \right\rangle )\left| {\phi (p)} \right\rangle },
\end{equation}
where $p$ is the eigenvalue of $\hat P$.

Postselection is then performed on the system and usually projected to a state $\left| {{\psi _f}} \right\rangle  = \frac{1}{{\sqrt 2 }}[\exp ( - i\rho )\left| H \right\rangle  - \exp (i\rho )\left| V \right\rangle ]$, which is nearly orthogonal to the preselected state. Here $\rho$ is the postselection angle. Postselection induces the meter state to collapse to an unnormalized redistribution as 
\begin{equation}
\label{eqS2}
\tag{S2}
D{(p)_{SWI}} = {\left| {\left\langle {\phi (p)} \right|\left\langle {{\psi _f}} \right|\left. {{\Psi _{SWI}}} \right\rangle } \right|^2}\\
 = \frac{{\Phi (p)}}{2}[1 - \cos (kp + 2\rho )].
\end{equation}
$\Phi (p) = {\left| {\left\langle {\phi (p)} \right.\left| {\phi (p)} \right\rangle } \right|^2}$ is the initial momentum spatial distribution. Postselection also leads to an observable postselection probability ${P_{SWI}}$ and weak value ${\langle {\hat A} \rangle _w}$ for the SWI
\begin{equation}
\label{eqS3}
\tag{S3}
{P_{SWI}} = {\left| {\left\langle {{\psi _f}} \right|\left. {{\Psi _{SWI}}} \right\rangle } \right|^2}\\
 = \frac{1}{2}[1 - {e^{ - \sigma _p^2{k^2}}}\cos (k{p_0} + 2\rho )]\\
 \approx {\sin ^2}(\rho ),
\end{equation}
\begin{equation}
\label{eqS4}
\tag{S4}
{\langle {\hat A} \rangle _w} = \frac{{\left\langle {{\psi _f}} \right|\hat A\left| {{\psi _i}} \right\rangle }}{{\left\langle {{\psi _f}} \right|\left. {{\psi _i}} \right\rangle }} = i\cot (\rho ).
\end{equation}
Often one can obtain $k$ from the shift of the mean value of the P pointer, which is calculated as follows
\begin{equation}
\label{eqS5}
\tag{S5}
\Delta {p_{SWI}} = \frac{{\int {pD{{(p)}_{SWI}}dp} }}{{\int {D{{(p)}_{SWI}}dp} }} - {p_0}\\
 = \frac{1}{{2{P_{SWI}}}}\sigma _p^2k{e^{ - \sigma _p^2{k^2}}}\sin (k{p_0} + 2\rho )\\
 \approx k\sigma _p^2{\mathop{\rm Im}\nolimits} ({\langle {\hat A} \rangle _w}).
\end{equation}
The approximation is feasible with $k{p_0}/2 \ll \rho  \ll 1$.

Next, we consider multiple weak interactions (MWI), where the MWI operator is denoted by [5,6]
\begin{equation}
\label{eqS6}
\tag{S6}
\hat \Gamma  = {\hat U^N} = {e^{i\frac{k}{2}N\hat A\hat P}}.
\end{equation}
MWI with $N = 1$ for the SWI. After MWI, these interactions allow the coupling system-meter state to evolve as
\begin{equation}
\label{eqS7}
\tag{S7}
\left| {{\Psi _{MWI}}} \right\rangle  = \int {dp\frac{1}{{\sqrt 2 }}({e^{iN\frac{{kp}}{2}}}\left| H \right\rangle  + {e^{ - iN\frac{{kp}}{2}}}\left| V \right\rangle )\left| {\phi (p)} \right\rangle } .
\end{equation}
With the same postselection as SWI, the distribution of the meter state is given by
\begin{equation}
\label{eqS8}
\tag{S8}
D{(p)_{MWI}} = \frac{{\Phi (p)}}{2}[1 - \cos (Npk + 2\rho )].
\end{equation}
The postselection probability ${P_{MWI}}$ and weak value ${\langle {\hat M} \rangle _w}$  corresponding to MWI are
\begin{equation}
\label{eqS9}
\tag{S9}
{P_{MWI}} = \frac{1}{2}[1 - {e^{ - \sigma _p^2{{(Nk)}^2}}}\cos (N{p_0}k + 2\rho )] \approx {\sin ^2}(\rho ),
\end{equation}
\begin{equation}
\label{eqS10}
\tag{S10}
{\langle {\hat M} \rangle _w} = {\langle {N\hat A} \rangle _w} = \frac{{\left\langle {{\psi _f}} \right|N\hat A\left| {{\psi _i}} \right\rangle }}{{\left\langle {{\psi _f}} \right|\left. {{\psi _i}} \right\rangle }} = iN\cot (\rho ).
\end{equation}
The shift of the mean value of the P pointer as
\begin{equation}
\label{eqS11}
\tag{S11}
\Delta {p_{MWI}} = \frac{1}{{2{P_{MWI}}}}\sigma _p^2Nk{e^{ - \sigma _p^2{{\left( {Nk} \right)}^2}}}\sin \left( {Nk{p_0} + 2\rho } \right)\\
 \approx k\sigma _p^2{\mathop{\rm Im}\nolimits} ( {{{\langle {N\hat A} \rangle }_w}} ).
\end{equation}
For the I pointer, let the initial intensity without postselection be ${I_{init}}$ and through the same evolution of the P pointer with SWI, the intensity detected after postselection is
\begin{equation}
\label{eqS12}
\tag{S12}
{I_{SWI}} = {I_{init}}{P_{{\rm{SWI}}}} = \frac{{{I_{init}}}}{2}[1 - {e^{ - \sigma _p^2{k^2}}}\cos ({p_0}k + 2\rho )].
\end{equation}
The shift of the I pointer with SWI is denoted as
\begin{equation}
\label{eqS13}
\tag{S13}
\Delta {\ell _{SWI}} = \frac{{\Delta {I_{SWI}}}}{{I_{SWI}^{k = 0}}} = \frac{{{I_{SWI}} - I_{SWI}^{k = 0}}}{{I_{SWI}^{k = 0}}}\\
 = \frac{{{e^{ - \sigma _p^2{k^2}}}2{p_0}k\sin (\rho) }}{{2{{\sin }^2}(\rho) }}\\
 \approx {e^{ - \sigma _p^2{k^2}}}{p_0}k{\mathop{\rm Im}\nolimits} ({\langle {\hat A} \rangle _w}).
\end{equation}
While for MWI, the postselection intensity evolves as
\begin{equation}
\label{eqS14}
\tag{S14}
{I_{MWI}} = {I_{init}}{P_{MWI}}\\
 = \frac{{{I_{init}}}}{2}[1 - {e^{ - \sigma _p^2{{(Nk)}^2}}}\cos (N{p_0}k + 2\rho )].
\end{equation}
The corresponding shift of I pointer with MWI is
\begin{equation}
\label{eqS15}
\tag{S15}
\Delta {\ell _{MWI}} = \frac{{\Delta {I_{MWI}}}}{{I_{MWI}^{k = 0}}}  \approx {e^{ - \sigma _p^2{{\left( {Nk} \right)}^2}}}{p_0}k{\mathop{\rm Im}\nolimits} ( {{{\langle {N\hat A} \rangle }_w}} ).
\end{equation}
From Eqs.~(\ref{eqS11}) and (\ref{eqS15}), it is evident that the shifts of the P and I pointers of MWI are amplified by a factor of $N$ compared to SWI, as elaborated in the main text.

\section{\label{sec2}Supplementary Section 2: Dual-pointer evolution and anomalous weak values high signal-to-noise detection}
\begin{figure}[htpb] 
\centering
\renewcommand\thefigure{S\arabic{figure}}
\includegraphics[width=0.7\linewidth]{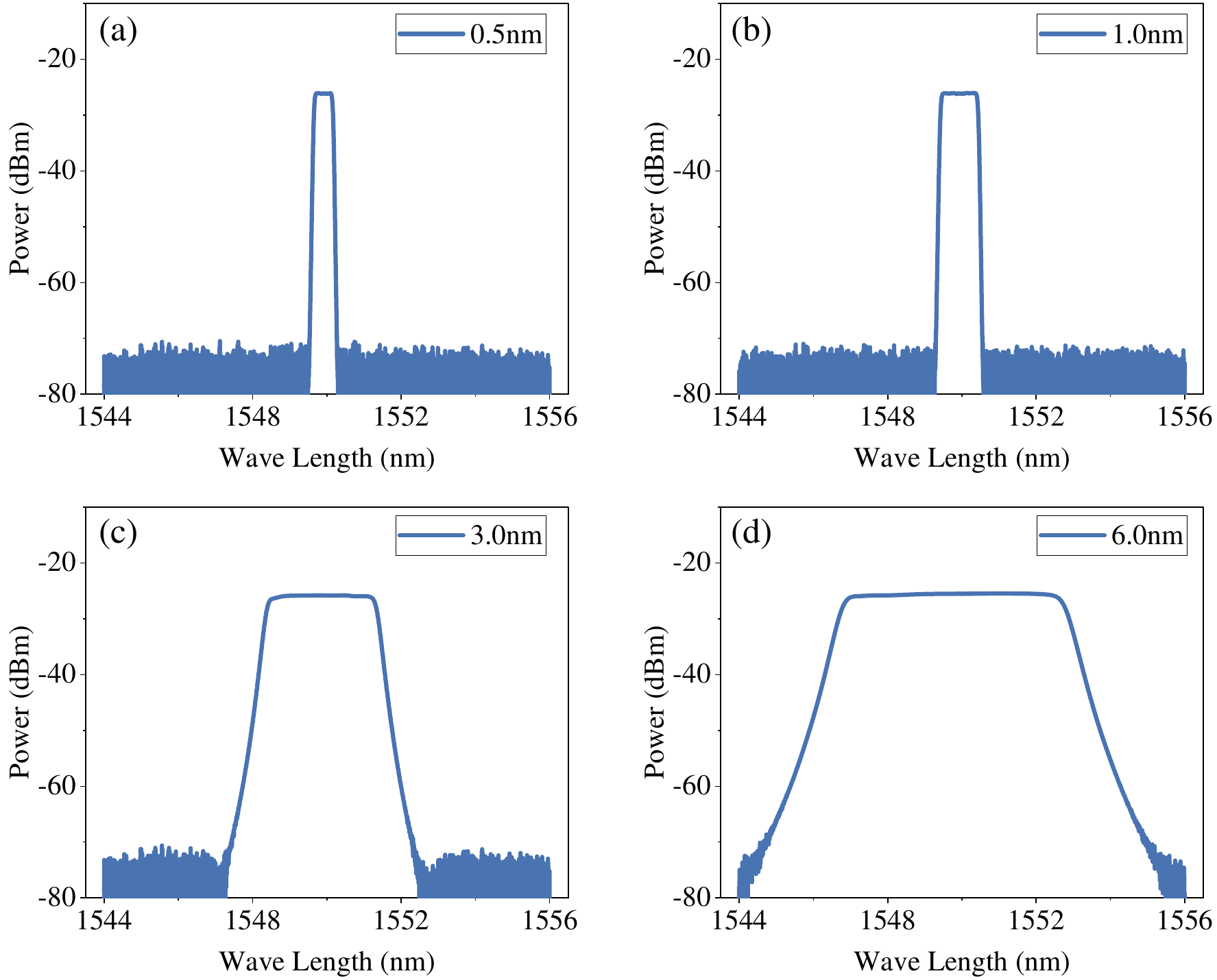}
\caption{Initial spectrograms of the variable spectrum noise source (VSNS) for four spectral widths 0.5 nm, 1.0 nm, 3.0 nm and 6.0 nm.}
\label{figS1}
\end{figure}

Due to the effect of the incident variable spectrum noise source (VSNS), we show the four different spectral widths of the VSNS used for the experiment, with spectral
widths of 0.5 nm, 1.0 nm, 3.0 nm, and 6.0 nm, espectively. The experimentally measured initial spectra are shown in Fig.~\ref{figS1}.  It can be seen that the top of the spectrum of each VSNS is flat, the center wavelength is 1550 nm, and the electronic noise of the spectrometer is about -75 dBm.

\begin{figure}[htpb] 
\centering
\renewcommand\thefigure{S\arabic{figure}}
\includegraphics[width=0.9\linewidth]{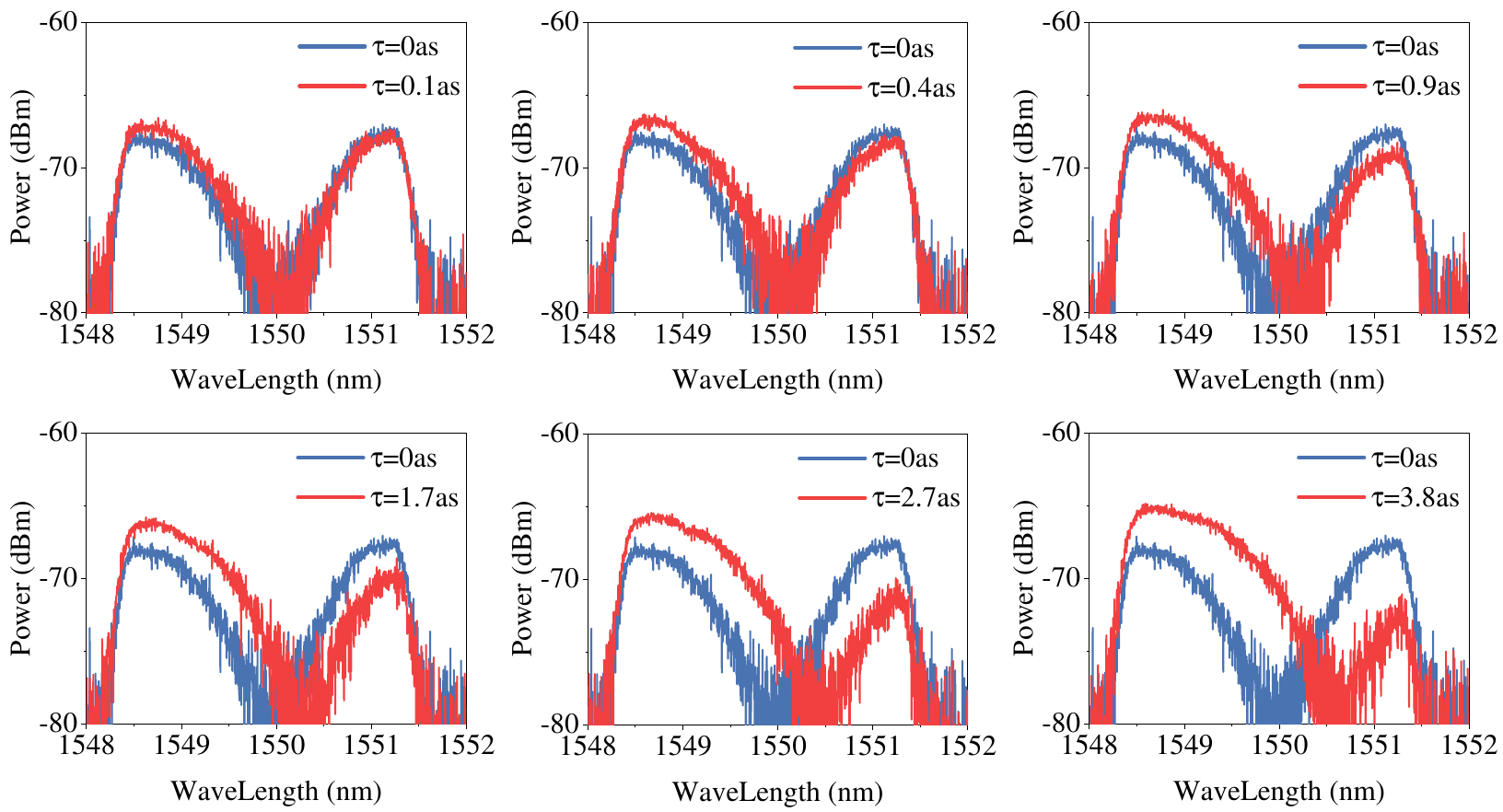}
\caption{The evolution of postselection spectrum with $\tau$ under $N = 1$. Compared to $\tau=0$, the postselection spectrum experiences both the P-pointer and the I-pointer shifts as $\tau$ increases.}
\label{figS2}
\end{figure}

Two key concepts, i.e. the P pointer and the I pointer, are analyzed in the main text to describe the evolution of the meter state after postselection from two distinct perspectives. For MWI with $N = 1$, the evolution is gauged employing a spectrometer under the postselection angle $\rho=0.002$ rad. Figure.~\ref{figS2} illustrates the partial evolution of the incident VSNS with a spectral width of 3.0 nm. The postselection spectrum initially displays bimodal distribution at $\tau=0$ as. The right peak gradually diminishes while the left peak becomes more prominent and eventually dominates as $\tau$ increases. Meanwhile, the overall intensity gradually increases and the spectrum shifts to the right. The horizontal shift of the postselection spectrum corresponds to the shift of the P pointer, and the vertical shift aligns with the I pointer shift.

\begin{figure}[htpb] 
\centering
\renewcommand\thefigure{S\arabic{figure}}
\includegraphics[width=0.8\linewidth]{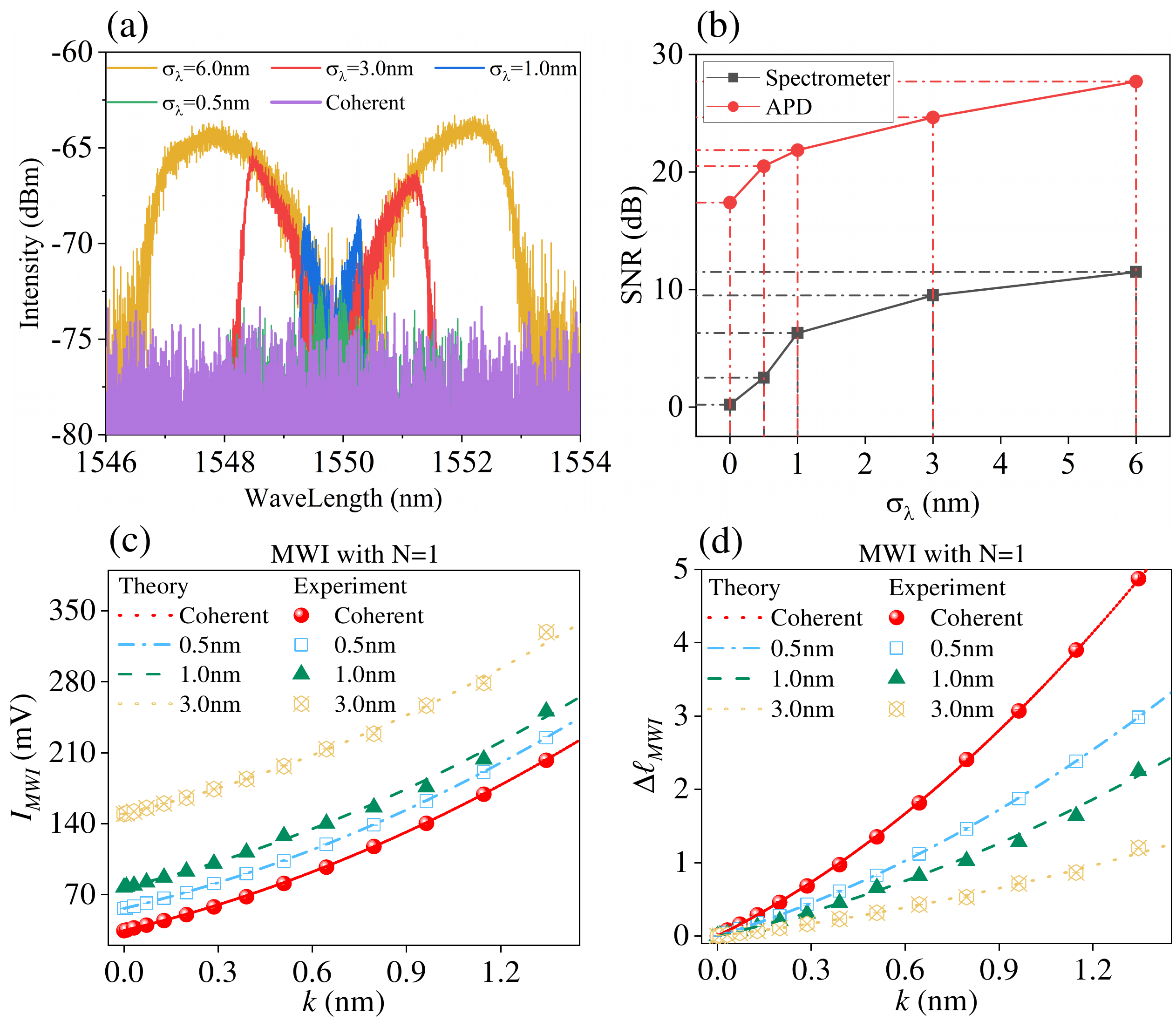}
\caption{(a) Bimodal spectrum distributions of various spectral-width VSNS after postselection. (b) Measured SNR by the spectrometer and APD after postselection. (c) and (d) For MWI with $N = 1$, intensity ${I_{MWI}}$ and shift $\Delta {\ell _{MWI}}$ detected by APD under various spectral-width sources versus displacement difference $k$. The error bars are the standard deviation of twenty repetitive experiments.}
\label{figS3}
\end{figure}

For an examination of the I pointer, we perform high-precision measurements of the displacement difference $k=c\tau$ using high signal-to-noise detection and narrow linewidth coherent light. According to Eq.~(\ref{eq4}) in the main text, the shift of I pointer is inversely proportional to the spectral width of incident light, contrasting with the P pointer shift. Figure~\ref{figS3} illustrates the measured results of the $k$ and signal-to-noise ratio (SNR) implementing the MWI with $N = 1$ for $\rho=0.002$ rad. The SNR defined as
\begin{equation}
\label{eq16}
\tag{S16}
\Re  = 10{\log _{10}}\frac{{{I_{MWI}}}}{{{I_{noise}}}},
\end{equation}
where ${I_{MWI}}=0.5$ mV denotes the mean electronic noise of the avalanche photodiode (APD). 

The spectral distributions and the SNR of coherent source with a linewidth of 400 kHz and VSNS after postselection are shown in Figs.~\ref{figS3}(a) and \ref{figS3}(b). The postselected SNR of the spectrometer decreases as the spectral width narrows, and the SNR of the 0.5 nm VWSN stands at only 2.5 dB, which is the narrowest spectral width distinguishing a bimodal distribution using the spectrometer. For a single-mode coherent source, the signal distribution is overwhelmed by the electrical noise, rendering its bimodal distribution indistinguishable. However, a substantial SNR of up to 17.5 dB is achieved using the APD to measure the postselected state of the coherent light. For MWI with $N = 1$, Figs.~\ref{figS3}(c) and \ref{figS3}(d) show the postselected intensity ${I_{MWI}}$ and shift $\Delta {\ell _{MWI}}$ of the I pointer detected by the APD varying with the $k$. It is observed that the results in Fig.~\ref{figS3}(d) validate the Eq.~(\ref{eq4}) in the main text, demonstrating an inverse relationship between the $\Delta {\ell _{MWI}}$ and the spectral width. The uncertainty of the intensity $\delta I$ for coherent source and VSNS with 0.5 nm, 1.0 nm, 3.0 nm are less than 0.045 mV, 0.072 mV, 0.11 mV and 0.21 mV, respectively. The corresponding precisions $\delta k$ are 497.8 $fm$, 782.7 $fm$, 1190.6 $fm$ and 2312.2 $fm$. It provides convincing evidence for the inverse relationship between the measurement precision of the I pointer and the spectral width of the incident light.

\begin{figure}[htpb] 
\centering
\renewcommand\thefigure{S\arabic{figure}}
\includegraphics[width=0.9\linewidth]{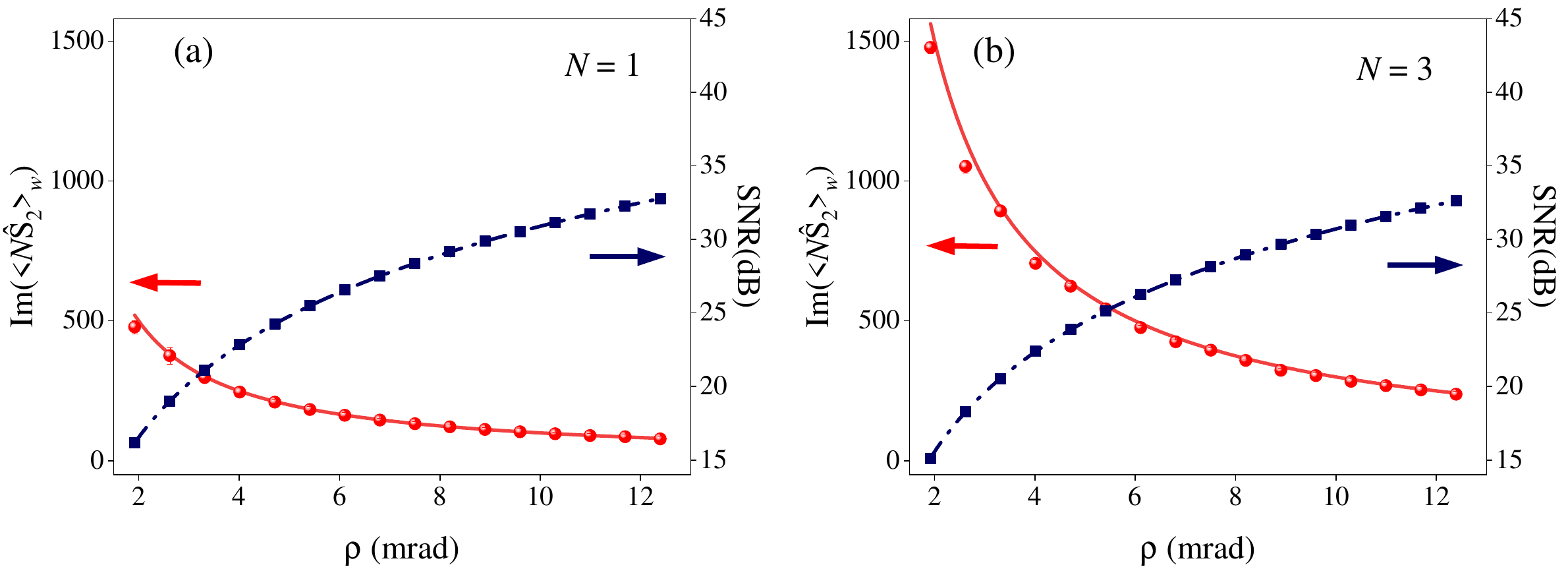}
\caption{Anomalous weak values ${{\mathop{\rm Im}\nolimits} ( {{{\langle {N{{\hat S}_2}} \rangle }_w}} )}$ and SNR versus $\rho$ with weak interactions (a) $N = 1$ and (b) $N = 3$. The dots are the experimental results, and the lines show the exact theoretical curves. The error bars are the standard deviation of twenty repetitive experiments.}
\label{figS4}
\end{figure}

We also achieve high signal-to-noise detection of anomalous weak values ${{\mathop{\rm Im}\nolimits} ( {{{\langle {N{{\hat S}_2}} \rangle }_w}} )}$, as depicted in Fig.~\ref{figS4}. By implementing triple weak interactions $N = 3$, we manage to capture extremely high weak values with a significant SNR. Specifically, an anomalous weak value 238 is obtained at SNR = 32.6 dB for postselection angle $\rho=0.0124$ rad. Decreasing $\rho$ further, with the SNR above 15 dB, we observe a significant anomalous weak value 1478. Such anomalous weak values reveal an intriguing phenomenon, demonstrating a deeper level of violation of the Leggett-Garg inequality (LGI), as shown in Fig.~\ref{fig6} of the main text.

\section{\label{sec3}Reference}
\noindent[1]	P. Yin, W. Zhang, L. Xu, Z. Liu, W. Zhuang, L. Chen, M. Gong, Y. Ma, X. Peng, G. Li, J. Xu, Z. Zhou, L. Zhang, G. Chen, C. F. Li, and G. C. Guo. Improving the precision of optical metrology by detecting fewer photons with biased weak measurement. Light Sci. Appl.  {\bf 10}, 103 (2021).

\noindent[2]	X. Y. Xu, Y. Kedem, K. Sun, L. Vaidman, C. F. Li, and G. C. Guo. Phase Estimation with Weak Measurement Using a White Light Source. Phys. Rev. Lett. {\bf 111}, 033604 (2013).

\noindent[3]	Z. Zhang, G. Chen, X. Xu, J. Tang, W. Zhang, Y. Han, C. F. Li, and G. C. Guo. Ultrasensitive biased weak measurement for longitudinal phase estimation. Phys. Rev. A {\bf 94}, 053843 (2016).

\noindent[4]	Y. Wang, W. Zhang, S. Chen, S. Wen, and H. Luo. Multiple-weak-value quantum measurement for precision estimation of time difference. Phys. Rev. A {\bf 105}, 033521 (2022).

\noindent[5]	F. Piacentini, A. Avella, M.P. Levi, M. Gramegna, G. Brida, I.P. Degiovanni, E. Cohen, R. Lussana, F. Villa, A. Tosi, F. Zappa, and M. Genovese. Measuring Incompatible Observables by Exploiting Sequential Weak Values. Phys. Rev. Lett. {\bf 117}, 170402 (2016).

\noindent[6]	Y. Kim, S. Y. Yoo, and Y. H. Kim. Heisenberg-Limited Metrology via Weak-Value Amplification without Using Entangled Resources. Phys. Rev. Lett. {\bf 128}, 040503 (2022).

\end{document}